\documentclass[conference,10pt,letterpaper]{IEEEtran}

\ifCLASSINFOpdf
\else
\fi

\usepackage{graphicx, bbm, latexsym, amsfonts, verbatim, amsmath, amssymb, algorithm, algorithmic, color, verbatim}
\usepackage{color, url}
\usepackage{tikz}
\usepackage{multirow}
\usepackage{array,booktabs}
\usepackage{epstopdf}
\usepackage{caption}
\usepackage{subcaption}

\newtheorem{corollary}{Corollary}
\newtheorem{define}{Definition}

\newtheorem{proposition}{Proposition}

\newtheorem{theorem}{Theorem}

\newcommand{\zred}{\textcolor{black}}
\newcommand{\tg}{\textcolor{black}}

\def\BEA{\begin{eqnarray}}
\def\EEA{\end{eqnarray}}

\def\BEAN{\begin{eqnarray*}}
\def\EEAN{\end{eqnarray*}}

\def\BE{\begin{equation}}
\def\EE{\end{equation}}

\begin{document}
%
\title{Survivable Probability of SDN-enabled Cloud Networking with Random Physical Link Failure}

\author{\IEEEauthorblockN{Zhili~Zhou}
\IEEEauthorblockA{United Airlines\\
Chicago, IL USA\\
Email: zhilizhou@gmail.com}
\and
\IEEEauthorblockN{Tachun~Lin}
\IEEEauthorblockA{Department of Computer Science and Information Systems\\
Bradley University, Peoria, IL USA\\
Email: djlin@bradley.edu}
}
\maketitle

\begin{abstract}
Software-driven cloud networking is a new paradigm in orchestrating physical resources (CPU, network bandwidth, energy, storage) allocated to network functions, services, and applications, which is commonly modeled as a cross-layer network. This model carries a physical network representing the physical infrastructure, a logical network showing demands, and logical-to-physical node/link mappings. In such networks, a single failure in the physical network may trigger cascading failures in the logical network and disable network services and connectivity. In this paper, we propose an evaluation metric, survivable probability, to evaluate the reliability of such networks under random physical link failure(s). We propose the concept of base protecting spanning tree and prove the necessary and sufficient conditions for its existence and relation to survivability. We then develop mathematical programming formulations for reliable cross-layer network routing design with the maximal reliable probability. Computation results demonstrate the viability of our approach.

\end{abstract}

\section{Introduction}
In a cloud computing architecture, hardware/physical infrastructures \tg{are physical resources composed of} datacenters and communication networks~\cite{develder2012optical}. \tg{All cloud services above are then realized} through cloud network mappings, \tg{which include} the mappings of demands (modeled as a logical network) onto the physical infrastructure and the deployment/allocation of virtual machines and physical resources\tg{~\cite{chowdhury2012vineyard}}.

Software-defined networking (SDN)~\cite{sezer2013we} and network function virtualization (NFV)~\cite{ETSI2015NFVinfrastructure}, which allow fine-grained control, orchestration, and management of heterogeneous physical resources, promise a future open marketplace where applications can be rapidly deployed, programmed, and operated on a converged infrastructure during their life cycles~\cite{kang2014software}. A software-defined infrastructure (SDI), such as SAVI~\cite{kang2014software}\cite{kang2013savi}, supports virtual network function (VNF) realization and enables application/service/infrastructure programmability.


The high-availability, reliability, and resilience of a software-defined infrastructure promises better cloud services, thus it becomes the key underlying network structure for SDI control and management~\cite{kreutz2015software}\cite{yang2015software}. The cloud networking facilitated with SDN/NFV is often modeled as a cross-layer network. In such a network, link or node failures in the SDN-enabled physical infrastructure may disrupt not only its own connectivity and capacity, but also fail or interrupt the users' demands for cloud applications and services in network abstraction (logical networks). Hence, leveraging SDN's programmability, our interest in this paper is to discover the optimal design of SDN-enabled cloud networking with less chances of failure.


The survivable network design in single-layer networks, which guarantees the network connectivity after network component's failure(s), was studied in~\cite{cholda2005network}\cite{heegaard2009network}\cite{ramamurthy1999survivable}. They proposed two mechanisms, namely protection and restoration, which were broadly adopted in modern networks. Nevertheless, the same mechanisms cannot be directly applied to the design of survivable cross-layer network, where the connectivity of the logical network remains after a single or multiple failures of physical network component. To overcome this challenge,~\cite{KurThi05}\cite{TodRam07}\cite{parandehgheibi2014survivable} studied mappings of logical links onto disjoint physical paths, and~\cite{ModNar01}\cite{lee2011cross}\cite{rahman2013svne} proposed the mappings avoiding cross-layer cuts. Another line of investigation is through route selection~\cite{parandehgheibi2014survivable}\cite{ye2014survivable}\cite{ye2015survivable}. While approaches in~\cite{ModNar01}\cite{lee2011cross} required the enumeration of all cutsets or path sets, Zhou et al.~\cite{zhou2017survivable}\cite{zhou2017novel} proposed the concept of protecting spanning tree set which avoided such enumeration. 

While most works in~\cite{KurThi05}--\cite{zhou2017survivable} are concentrated on finding a survivable cross-layer mapping, an NP-hard problem, it is still an open question on which routings should be chosen when a survivable mapping does not exist. Yallouz et al.~\cite{yallouz2014tunable}\cite{yallouz2017tunable} introduced the \emph{level of survivability} which provides ``a quantitative measure for specifying any desired level of survivability'' under a single-layer network setting through spanning trees. Zhou et al.~\cite{zhou2017novel}\cite{zhou2012} also introduced a similar concept of \emph{partial survivability} in cross-layer networks. Based on these works, we define the \textbf{survivable probability} of a given cross-layer network, a new evaluation metric which is the probability of the logical network to remain connected after any physical link failure. We wish to point out that \emph{survivable probability} is evaluated when each physical link may fail with a uniform or random failure probability, while the physical link failure probability considered in~\cite{zhou2017novel}\cite{zhou2012} is always 100\%. Also, the survivable probability discussed here is for cross-layer networks instead of single-layer networks discussed in~\cite{yallouz2014tunable}\cite{yallouz2017tunable}.

Another related work~\cite{lee2014maximizing} also studied the cross-layer network reliability by calculating the failure polynomials under random physical link failure. Our proposed study is different from~\cite{lee2014maximizing} in two aspects: (1) the solution approach in~\cite{lee2014maximizing} is based on cross-layer cutsets, while our approach eliminates the needs to enumerate all cutsets, and (2) we consider not only uniform but also random failure probability on all physical links.


Our contributions in this paper are:
(1) we define a new evaluation metric of reliability, i.e., the survivable probability, for cross-layer networks; instead of approximating with failure polynomials, we propose an exact method to identify a network's survivable probability;
(2) we introduce the concept of \emph{base protecting spanning tree set} and prove its existence;
(3) we prove that the survivable probability of a network with unified physical link failure probability is equivalent to the cross-layer network design with the minimal number of shared physical links in a protecting spanning tree set;
(4) compared with the failure polynomial approximation for the reliability of cross-layer networks, our approach avoids the enumeration of all cross-layer cutsets and requires at most $|E_P|$ protecting spanning trees;
(5) last but not least, the proposed base protecting spanning tree set enables SDI control and management and can be used to improve resilience of cloud services under unpredictable failure(s), which can be treated as an auxiliary protection scheme for SDN without introducing redundant physical resources.

The rest of this paper is organized as follows: Section~\ref{sec:problem} provides formal definitions and problem description for survivable probability and protecting spanning trees. Solution approaches for survivable probability are presented in Section~\ref{sec:approach}. Section~\ref{sec:result} shows the simulation results for survivable probability based on the protecting spanning tree set in a cross-layer network, followed by the conclusion in Section~\ref{sec:conclusion}.

\section{Definitions and Problem Description}\label{sec:problem}
Given a cloud network with a physical network $G_P=(V_P, E_P)$ and a logical network $G_L=(V_L,E_L)$, where each logical node has an one-to-one mapping onto a physical node (denoted as $m(s)=i$ with $s\in V_L$ and $i\in V_P$), and each logical link is mapped one-to-one onto a physical path (denoted as $m(u)=p_u$ with $u\in E_L$ and $p_u \subset E_P$). We let $\mathcal{M}(G_L,G_P)=\{m(s), m(u): s\in V_L, u\in E_L\}$ represent the mapping between logical to physical networks. Notations and parameters used in this paper are listed in Table~\ref{tbl:notation}.
\begin{table}[t]
\begin{tabular}{p{2cm}|p{6cm}}
\hline\hline
 \rule{0pt}{9pt}Notation       &Description\\
\hline
 \rule{0pt}{8pt} $G_P = (V_P,E_P)$ &Physical network with node set $V_P$ and link set $E_P$. $i,j$ are node indices and $e$ is the link index\\
 \rule{0pt}{8pt} $G_L = (V_L,E_L)$ &Logical network with node set $V_L$ and link set $E_L$. $s,t$ are node indices and $u, v$ are link indices\\
 \rule{0pt}{8pt} $\mathcal{M}(G_L, G_P)$ &The mapping between physical and logical networks, including node and link mappings \\
 \rule{0pt}{8pt}$\mathcal{P}_{u}$ & The set of routings (physical paths) for $u\in E_L$, where $p_u\in \mathcal{P}_u$ is its element\\
 \rule{0pt}{8pt} $\lambda$      & $\lambda = [\tau, \{p_{u}: u\in \tau\}]$,  where $\tau\subset E_L$ is a logical spanning tree and $\{p_u: u\in \tau\}$ is the routings of the tree branches\\
 \rule{0pt}{8pt} $\mathcal{T}$  &A spanning tree set, where $\lambda \in \mathcal{T}$\\
 \rule{0pt}{8pt} $E_{P}(\lambda)$ &All physical links utilized by the routings of $\lambda$'s branches\\
 \rule{0pt}{8pt} $E^{C}_{P}(\mathcal{T})$ & Common physical links shared by the routings of all $\lambda\in \mathcal{T}$\\
 \rule{0pt}{8pt} $M(E_L)$       &A set of logical link mappings, where $m(E_L)\in M(E_L)$ is one instance of logical link mapping\\
 \rule{0pt}{8pt} $R(m(E_L))$    &A set of physical links whose failures will disconnect $G_L$ for a given mapping $m(E_L)$\\
 \rule{0pt}{8pt} $\Phi(G_P,G_L)$&The maximal survivable probability of a cross-layer network\\
\hline
\hline
 \rule{0pt}{9pt}Parameter&Description\\
\hline
 \rule{0pt}{8pt} $\rho_e$       &Physical link failure probability with $e\in E_P$\\
 \rule{0pt}{8pt} $\rho$         &Unified physical link failure probability with $e\in E_P$\\
\hline\hline
\end{tabular}
\vspace{2pt}
\caption{Notations and parameters}
\label{tbl:notation}
\end{table}

\begin{define}\label{def:netSp}
Given $G_P$, $G_L$, and failure probability of physical link $\rho_e, e\in E_P$, the \textbf{survivable probability} of a cloud network is the probability of the logical network to remain connected after physical link failure(s).
\end{define}

\textbf{Problem Description:} Given a cloud network $G_P, G_L$, and $\rho_{e}, e\in E_P$. We wish to identify its mappings $\mathcal{M}(G_L, G_P)$ which provide the maximal survivable probability.

\begin{define}\label{def:proTree}
Given $G_P$, $G_L$, a logical spanning tree $\tau \in G_L$ and its routings $\{p_u: u\in \tau\}$. $\tau$ is a \textbf{protecting spanning tree} if it remains connected after one or multiple physical link failures.
\end{define}
We let $\lambda=[\tau, \{p_u: u\in \tau\}]$ denote a protecting spanning tree in a cloud network and $E_{P}(\lambda)=\{e: e\in \cup_{u\in \tau}p_u\}$ be the physical link set utilized by the routings of $\tau$.
With the introduction of physical link failure probability, we may further define the maximal protecting spanning tree as follows.
Given cross-layer network and 
any logical link mapping, a \textbf{maximal protecting spanning tree} is a protecting spanning tree with the maximal survivable probability.
\begin{define}\label{def:proTreeSp}
Given $G_P$, $G_L$, $\rho_e, e\in E_P$, the \textbf{survivable probability of} \textbf{$\lambda$}$=[\tau, \{p_u: u\in \tau\}]$ is $\text{Prob}(\lambda)=\prod_{e\in E_{P}(\lambda)}(1-\rho_e)$.
\end{define}
This definition is derived from the fact that $\tau$ will remain connected only if all physical links utilized by the routings of $\tau$ are not disconnected.

\begin{figure}
\centering
\includegraphics[scale=0.6]{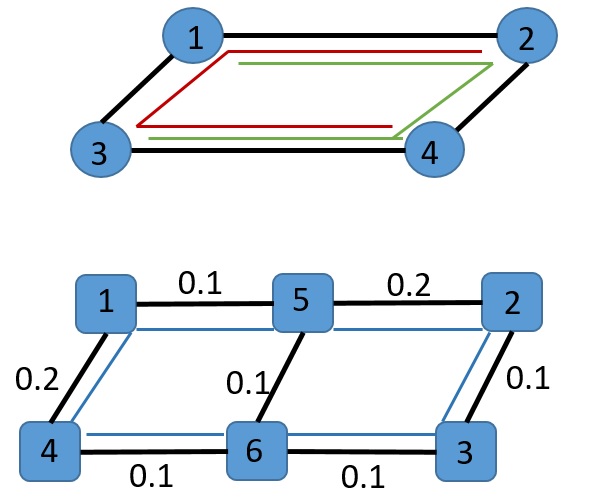}
\caption{Survivable probability with multiple protecting spanning trees}\label{fig:2tree}
\end{figure}

We let $\mathcal{T}=\{\lambda_i\}$ be a set of protecting spanning trees and $E^{C}_{P}(\mathcal{T})=\cap_{\lambda_i\in \mathcal{T}}E^{C}_{P}(\lambda_i)$ be the common physical links utilized by the routings of $\lambda_i\in \mathcal{T}$. We use Fig.~\ref{fig:2tree} as an instance to illustrate the concept of protecting spanning tree set and how it can be use to improve survivable probability. Given $G_L$ (top), $G_P$ (bottom), and $\rho_e, e\in E_P$ (noted on each physical link). We select a set of two protecting spanning trees: (red tree) $\lambda_1=[\{(2,1),(1,3),(3,4)\}, \{\{(1,5),(5,2)\}, \{(1,4),(4,6),(6,3)\}$, $\{(4,6),(6,3)\}\}]$, where  $E_{P}(\lambda_1)=\{(1,4)(1,5),(2,5),(3,6),(4,6)\}$ and the survivable probability of $\lambda_1$ is $(1-0.2)(1-0.1)(1-0.2)(1-0.1)(1-0.1)=0.46656$; (green tree) $\lambda_2 = [\{(1,2),(2,4),(4,3)\}, \{\{(1,5),(5,2)\},\{(2,3),(3,6),(6,4)\}$, $\{(4,6),(6,3)\}\}]$, where $V_P(\lambda_2)=\{(1,5),(2,3),(2,5),(3,6),(4,6)\}$ and the survivable probability of $\lambda_2=(1-0.1)(1-0.2)(1-0.1)(1-0.1)(1-0.1)=0.52488$. When considering $\lambda_1$ and $\lambda_2$ together, the common physical links used by the routings of both trees are $E^{C}_{P}(\lambda_1, \lambda_2)=\{(1,5),(2,5),(3,6),(4,6)\}$. Therefore, any failure(s) among these links would cause the failure of both $\lambda_1$ and $\lambda_2$. Hence, the survivable probability of $[\lambda_1,\lambda_2]= (1-0.1)(1-0.2)(1-0.1)(1-0.1) = 0.5832$ which is higher than that of either $\lambda_1$ or $\lambda_2$.

Derived from the example above, we have the following definition.
\begin{define}\label{def:treeSetSp}
Given $G_P, G_L, \rho_e, e\in E_P$, and a protecting spanning tree set $\mathcal{T}=\{\lambda_i\}$, the \textbf{survivable probability of $\mathcal{T}$} is $\text{Prob}(\mathcal{T})=\prod_{e\in E^{C}_{P}(\mathcal{T})}(1- \rho_{e})$.
\end{define}
\begin{define}\label{def:baseTreeSet}
Given $G_P$ and $G_L$, a \textbf{base protecting spanning tree set} for a cloud network has the same survivable probability as that of the cloud network.
\end{define}

\subsection{Survivable Probability, Link Mapping, and Base Protecting Spanning Tree Set}\label{subsec:netTreeSetRlt}
In this section, we explore the relationship among cross-layer networks, logical link mappings, and protecting spanning tree sets. We then identify a cross-layer network's survivable probability based on the base protecting spanning tree set.

Given a cross-layer network $G_P$, $G_L$, and cross-layer link mappings $m(E_L)=\{p_u, p_u\in \mathcal{P}_u, u\in E_L\}$ with $M(E_L)=\{m(E_L)\}$. We define and analyze the \textbf{maximal survivable probability of a cross-layer network}, $\Phi(G_P, G_L)$, as follows.
\begin{proposition}\label{prop:cldNetSurProb}
For a given cloud network $G_P$, $G_L$, and its mappings $M(E_L)$, $\Phi(G_P, G_L) = \max_{m(E_L)\in M(E_L)}\prod_{e\in R(m(E_L))}(1-\rho_e)$, where $R(m(E_L))$ denotes a set of physical links whose failure(s) disconnect $G_L$.
\end{proposition}
\begin{IEEEproof}
With Definition~\ref{def:netSp}, survivable probability of a cloud network is determined by physical links whose failures disconnect the logical network. As the mapping of logical links decides such physical link set, with logical link mapping set $M(V_L)$, $\Phi(G_P, G_L)=\max_{m(E_L)\in M(E_L)}\prod_{e\in R(m(E_L))}(1-\rho_e)$.
\end{IEEEproof}
We let $m^{*}(E_L)$ denote the logical link mapping with the maximal survivable probability, i.e., $m^{*}(E_L)=\arg_{m(E_L)\in M(E_L)}\Phi(G_P, G_L)$.
Next, we prove the existence of the base protecting spanning tree set with given $G_P$ and $G_L$.
\begin{proposition}\label{prop:existTreeSet}
Given $G_P$, $G_L$, and $M(E_L)$, the base protecting spanning tree set exists.
\end{proposition}
\begin{IEEEproof}
We prove this through two claims:
Claim 1: given a logical link mapping $m(E_L)\in M(E_L)$, a physical link is \emph{protected} by at least one protecting spanning tree if its failure does not disconnect $G_L$;
Claim 2: with the same mapping $m(E_L)$, a physical link is not protected by any protecting spanning tree if its failure causes the disconnection of $G_L$.

Proof of Claim 1: for $\lambda=[\tau, \{p_u: u\in \tau\}]$, any single or multiple link failures in $E_P\setminus E_{P}(\lambda)$ do not disconnect $\tau$. Hence, $\lambda$ guarantees the connectivity of $G_L$ after the failure(s) of link(s) in $E_P\setminus E_{P}(\lambda)$.

Proof of Claim 2: proof by contradiction. If the failure of a physical link causing the disconnection of the logical network can be protected by a protecting spanning tree, with Claim 1, the logical network should remain connected after its failure. Contradiction!

With Claims 1 and 2, let a tuple $[\bar{\mathcal{T}}, m(E_L)]$ represent a protecting spanning tree set and its mappings, where $\bar{\mathcal{T}}$ contains all protecting spanning trees and each tree protects at least one physical link. $R(m(E_L))$ is a set of physical links not protected by trees in $[\bar{\mathcal{T}}, m(E_L)]$. There exists a mapping $m'(E_L) \in M(E_L)$ such that $m'(E_L)=\arg_{m(E_L)\in M(E_L)}\max \prod_{e\in E^{C}_{P}([\bar{\mathcal{T}}, m(E_L)])}(1- \rho_{e})$. With Claim 2, $R(m'(E_L))=E^{C}_{P}([\bar{\mathcal{T}}, m'(E_L)])$. Therefore, with $m^{*}(E_L)$ as a logical link mapping, there exists a protecting spanning tree set which has the same survivable probability as that of the given cross-layer network. The conclusion holds.
\end{IEEEproof}
We may derive from Proposition~\ref{prop:existTreeSet} the necessary and sufficient conditions of a base protecting spanning tree set in a cross-layer network as follows.
\begin{theorem}\label{thm:ncSufBaseProTreeSet}
A protecting spanning tree set is a base protecting spanning tree set if and only if it provides the maximal survivable probability
$\prod_{e\in E^{C}_{P}(\mathcal{T})}(1- p_{e})$.
\end{theorem}
\begin{IEEEproof}
Proof of the necessary condition: given in Proposition~\ref{prop:existTreeSet}.\\
Proof of the sufficient condition by contradiction to Definition~\ref{def:baseTreeSet}: if there exists a protecting spanning tree set with higher survivable probability than that of the base protecting spanning tree set, it leads to higher cross-layer survivable probability. Contradiction!
\end{IEEEproof}
\begin{corollary}\label{col:survivable}
A survivable cross-layer network has 100\% survivable probability against arbitrary physical link failure.
\end{corollary}
\begin{IEEEproof}
With Theorem 1 in~\cite{zhou2017novel}, no common physical links are utilized by a protecting spanning tree set $\mathcal{T}$, i.e., $\bigcap_{u\in \cup_{\lambda\in \mathcal{T}}}p_u=\emptyset$. With Definition~\ref{def:netSp}, its survivable probability is 100\%.
\end{IEEEproof}
Proposition~\ref{col:survivable} indicates that survivable cross-layer networks are a subset of cross-layer networks with guaranteed 100\% survivable probability.

\subsection{Special Case: Unified Physical Link Failure Probability}\label{subsec:failProb}
Unified physical link failure probability considered in this section is a special case where the failure probability for all physical links is $\rho$.
We first study the maximal protecting spanning tree and have the following conclusion.
\begin{proposition}~\label{prop:mxTreeSpCase}
Given $G_P$, $G_L$, and physical link failure probability $\rho_e=\rho$ with $e\in E_P$, the maximal protecting spanning tree is a protecting spanning tree with the minimal number of physical links utilized in its branches' routes.
\end{proposition}
\begin{IEEEproof}
Based on Definition~\ref{def:proTreeSp}, the maximal  survivable probability  $= \max \prod_{e\in E_{P}(\lambda)}(1-\rho_e)$ $ = \max (1-\rho)^{|E_{P}(\lambda)|}$  (with unified physical link failure probability $\rho$), which is equivalent to $\min |E_{P}(\lambda)|$.
\end{IEEEproof}

\begin{theorem}~\label{thm:failProbMinLk}
Given $G_P$, $G_L$, and unified failure probability $\rho$, a base protecting spanning tree set with  $\min|E^{C}_{P}(\mathcal{T})|$ has survivable probability  $(1-\rho)^{\min|E^{C}_{P}(\mathcal{T})|}$.
\end{theorem}
\begin{IEEEproof}
Derived from Proposition~\ref{prop:mxTreeSpCase}, given  unified failure probability $\rho$, $G_L$ is disconnected if one or multiple physical links commonly used by the routings of $\lambda\in \mathcal{T}$ fail. Thus, $\min|E^{C}_{P}(\mathcal{T})|$ also leads to the maximal survivable probability. Therefore, the conclusion holds.
\end{IEEEproof}
Based on Theorem~\ref{thm:failProbMinLk}, the estimation of cross-layer network survivability with unified failure probability is equivalent to solving a cross-layer network design problem targeting the minimal number of shared physical links in its logical link mappings. The above proof also demonstrates that a base protecting spanning tree set and its mappings can provide a (partially) survivable network design along with a more precise evaluation metric on its survivability.

Compared with the cross-layer network reliability problem considered in~\cite{lee2014maximizing} where the reliability is approximated through enumerating exponential number of cross-layer cutsets and evaluating failure polynomials, the base protecting spanning tree set can provide exact computation for survivable probability under both uniform and random physical link failure probability. Theorem~\ref{thm:failProbMinLk} further demonstrates that the exact survivable probability problem in cross-layer networks is equivalent to the cross-layer network routing problem which guarantees minimal number of shared physical links by protecting spanning trees in a base protecting spanning tree set.

We wish to point out that with random physical link failure probability, the minimal cardinality of a set of physical links whose failures disconnect the logical network may not lead to the same survivable probability of the original cross-layer network.

\section{Solution Approach}\label{sec:approach}
Based on Theorems~\ref{thm:ncSufBaseProTreeSet} and~\ref{thm:failProbMinLk}, we present in this section the mathematical programming formulations to compute survivable probability of cross-layer networks and its related problems. We first present the formulation with unified physical link failure probability as a special case in Section~\ref{subsec:speCase}. Then, the generalized formulation with random physical link failure probability is presented in Section~\ref{subsec:baseTreeSet}.
Variables and parameters used in the formulations are listed in Table~\ref{tbl:vrPr}.
\begin{table}
\begin{tabular}{p{2cm}|p{6cm}}
\hline\hline
 \rule{0pt}{9pt} Variable &Description\\
\hline
\rule{0pt}{8pt} $x_{ij}$& Binary variable indicating whether $(i,j)$'s failure disconnects the logical network. If yes, $x_{ij}=1$; otherwise, $x_{ij}=0$\\
 \rule{0pt}{8pt} $y^{st}_{ij}$& Binary variable indicating whether logical link $(s,t)$ is routed through physical link $(i,j)$ or not. If yes, $y^{st}_{ij}=1$; otherwise, $y^{st}_{ij}=0$\\
 \rule{0pt}{8pt} $z_{st}$& Binary variable indicating whether logical link $(s,t)$ is connected and forms a protecting spanning tree. If yes, $z_{st}=1$; otherwise, $z_{st}=0$\\
 \rule{0pt}{8pt} $w^{ij}_{st}$& Binary variable indicating whether logical link $(s,t)$ is connected and forms a protecting spanning tree after physical link $(i,j)$ fails. If yes, $w^{ij}_{st}=1$; otherwise, $w^{ij}_{st}=0$\\
 \rule{0pt}{8pt} $g_{ij}$& Binary variable indicating whether physical link $(i,j)$ is shared by trees in a base protecting spanning tree set. If yes, $g_{ij}=1$; otherwise, $g_{ij}=0$\\ \hline\hline
 \rule{0pt}{9pt} Parameter & Description\\
\hline
 \rule{0pt}{8pt} $c_{ij}$  & The coefficient for physical link. With unified failure probability, $c_{ij}=1$; with random physical link failure probability, $c_{ij}=\ln(1-\rho_{ij})$\\
\hline\hline
\end{tabular}
\vspace{2pt}
\caption{Variables and parameters used in mathematical formulations}
\label{tbl:vrPr}
\end{table}

\subsection{Special Case: Survivable Probability of Cross-layer Networks with Unified Physical Link Failure Probability}\label{subsec:speCase}
We first present a mixed-integer programming formulation to generate the maximal protecting spanning tree, followed by a formulation to generate a base protecting spanning tree set for cross-layer network survivable probability.
\subsubsection{Maximal Protecting Spanning Tree}\label{subsubsec:maxTree}
Given unified physical link failure probability $\rho$, \emph{MaxPrctTree} is  a mixed integer programming formulation  with the objective to minimize the number of physical links utilized by tree branches' routings (based on Proposition~\ref{prop:mxTreeSpCase}).

\begin{align}
& \min_{x,y,z}\sum_{(i,j)\in E_P}x_{ij}\nonumber\\
s.t.    &\sum_{(i,j)\in E_P}y^{st}_{ij}-\sum_{(j,i)\in E_P}y^{st}_{ji}=
\left\{\begin{matrix}
z_{st}, &\,\mbox{if } i=s,\\
-z_{st}, &\,\mbox{if } i=t,\\
0, &\,\mbox{if } i\neq \{s, t\},
&\end{matrix}\right.\label{fm:lightpath1}\\
        &y^{st}_{ij}+y^{st}_{ji}  \leq x_{ij}, (s,t)\in E_L, (i,j)\in E_P\label{fm:minLinkSel}\\
        &\sum_{(s,t)\in E_L}z_{st} - \sum_{(t,s)\in E_L} z_{ts} =
\left\{\begin{matrix}
|V_L|-1, \text{if} s=s_0\\
-1,      \text{if} s\neq s_0, s\in V_L
&\end{matrix}\right.\label{fm:maxTreeTree1}\\
        &\sum_{(s,t) \in E_L}z_{st} =|V_L|-1, (i,j)\in E_P \label{fm:zBnForced}\\
        &x_{ij}, y^{st}_{ij}, z_{st} \in \{0,1\}, (s,t)\in E_L, (i,j)\in E_P\label{fm:maxTreeRegion}
\end{align}
Constraint (\ref{fm:lightpath1}) maps logical links onto physical paths and forms a logical spanning tree, in which $z_{st}$ on the right hand side indicates whether $(s,t)$ is a branch of a logical spanning tree or not.
Constraint (\ref{fm:minLinkSel}) indicates which physical links are utilized by the routings of a selected protecting spanning tree.
Constraints (\ref{fm:maxTreeTree1}) and (\ref{fm:zBnForced}) form a protecting spanning tree corresponding to the logical link mapping generated by constraint (\ref{fm:lightpath1}). Constraint~(\ref{fm:maxTreeRegion}) provides the feasible regions for all decision variables.

\subsubsection{Base Protecting Spanning Tree Set}\label{subsubsec:treeSet}
Extended from the \emph{MaxPrctTree} formulation, we present the mixed-integer programming formulation \emph{BasePrctTree} to compute a cross-layer network's survivable probability with unified physical link failure probability.
With Theorems~\ref{thm:ncSufBaseProTreeSet} and~\ref{thm:failProbMinLk}, the mixed-integer programming formulation has the objective to minimize the total number of physical links shared by protecting spanning trees.
\begin{align}
\min_{y,g} &\sum_{(i,j)\in E_P} g_{ij} \nonumber\\\
s.t. &\sum_{(i,j)\in E_P}y^{st}_{ij}-\sum_{(j,i)\in E_P}y^{st}_{ji}=
\left\{\begin{matrix}
\zred{1}, &\,\mbox{if } i=s,\\
\zred{-1}, &\,\mbox{if } i=t,\\
0, &\,\mbox{if } i\neq \{s, t\},
&\end{matrix}\right.\label{fm:lightpath2}\\
       &w^{ij}_{st}\leq 1-(y^{st}_{ij}+y^{st}_{ji}), (s,t)\in E_L, (i,j)\in E_P \label{fm:protTreeSet2}\\
       \sum_{(s,t)\in E_L}&w^{ij}_{st} - \sum_{(t,s)\in E_L} w^{ij}_{ts} =
\left\{\begin{matrix}
\zred{(1 - g_{ij})}, \qquad\text{if } s=s_0\\
\zred{(g_{ij}-1)/(|V_L|-1)},  \\
\qquad\quad\text{if } s\neq s_0, s\in V_L
&\end{matrix}\right.\label{fm:maxTreeTreeSet2}\\
       y^{st}_{ij}, &g_{ij} \in \{0,1\}, w^{ij}_{st}\geq 0, (s,t)\in E_L, (i,j)\in E_P\label{fm:maxTreeRegion2}
\end{align}
Similar to constraint (\ref{fm:lightpath1}), constraint (\ref{fm:lightpath2}) generates physical paths for links of tree branches of a spanning tree in a base protecting spanning tree set. \zred{Constraints (\ref{fm:protTreeSet2})--(\ref{fm:maxTreeTreeSet2}) generate a protecting spanning tree after any physical link's failure if the physical link is protected; otherwise, the physical link is identified as unprotected. With the information of unprotected physical links, the generated protecting spanning trees which protect at least one physical link are then identified as the base protecting spanning tree set.} Constraint~(\ref{fm:maxTreeRegion2}) provides the feasible regions for all decision variables.

\subsection{Survivable Probability of Cross-Layer Networks with Random Physical Link Failure Probability}\label{subsec:baseTreeSet}
In this section, we discuss a more generalized and realistic physical link failure scenario with random failure probability on physical links.
With Definition~\ref{def:baseTreeSet}, the objective used to select a base protecting spanning tree set is as follows:
\begin{align}
     &\max_{\tau, \mathcal{P}}\prod_{_{e\in \cup{u\in \tau}} p_{u}}(1-\rho_e)\label{obj:nonlinear}
\end{align}
The objective (\ref{obj:nonlinear}) is nonlinear. Applying the $\ln$ function on the objective converts it into linear form as follows:
\begin{align}
\max_{\tau, \mathcal{P}}\sum_{_{e\in \cup{u\in \tau}}p_u}\ln(1-\rho_e).\nonumber
\end{align}
The generalized formulation for cross-layer survivable probability is then
\begin{align}
\min_{y,g}&\sum_{(i,j)\in E_P}c_{ij}g_{ij}.\nonumber\\
s.t. & \text{ Constraints (\ref{fm:lightpath2}) -- (\ref{fm:maxTreeRegion2})}.
\end{align}
Here with unified physical link failure probability, $c_{ij}=1$;
and with random physical link failure probability, $c_{ij}=-\ln(1-p_{ij})$. Note here that since $p_{ij}$ and $1-p_{ij}$ are in (0,1] for all $(i,j)\in E_P$, and $\ln(1-p_{ij})$ is in $(-\infty, 0)$. Hence, maximizing negative $ln(1-p_{ij})$ equals to minimizing positive $c_{ij}$.

\section{Simulation Study}\label{sec:result}
In this section, we present our simulation settings and test cases, experiment design, and simulation results. The goal is to validate and demonstrate the efficiency of the proposed base protecting spanning tree set for a cloud network's survivable probability.
\subsection{Simulation Setup}\label{subsec:testCase}
\begin{figure}
\centering
\includegraphics[scale=0.43]{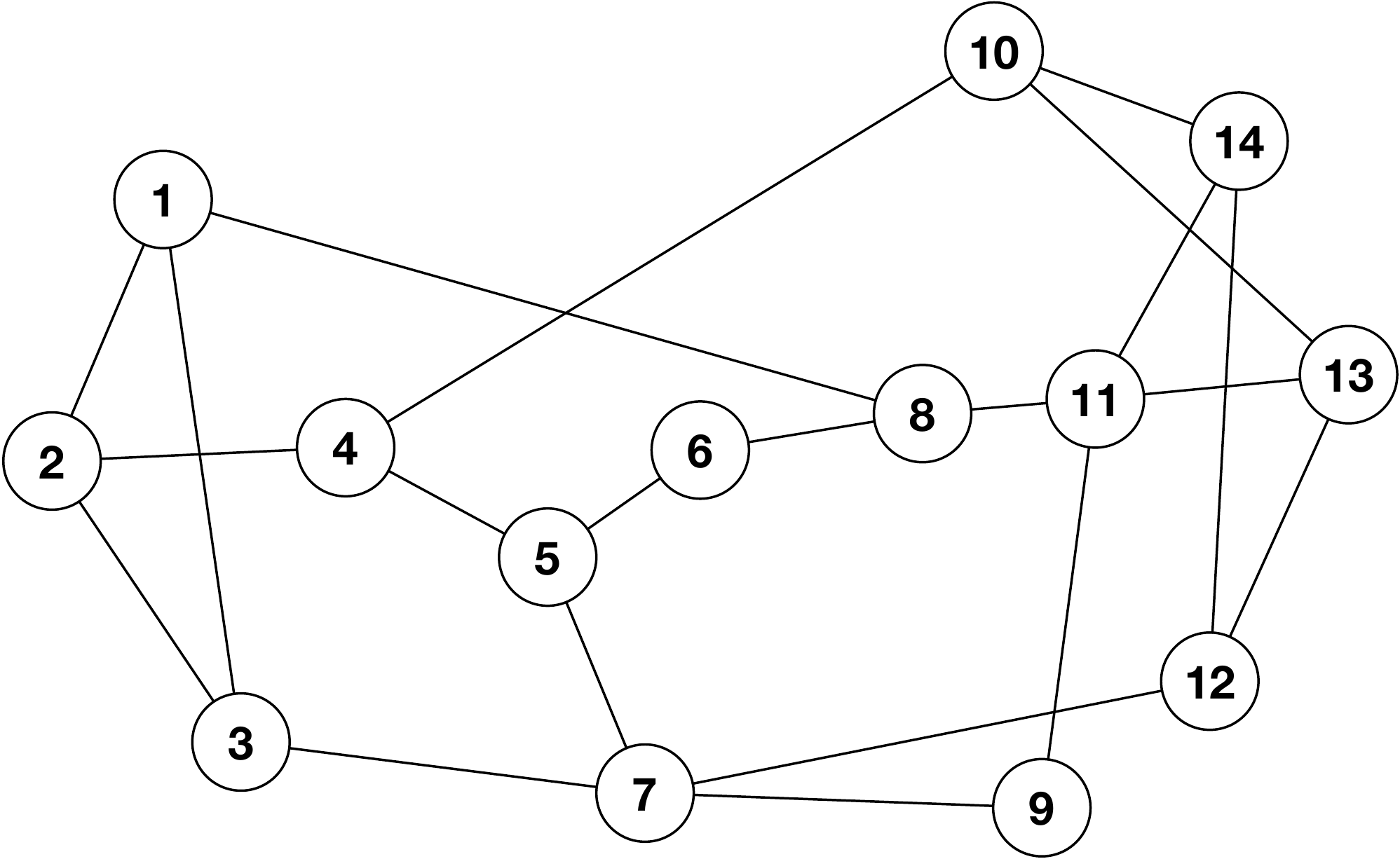}
\caption{NSF}
\label{fig:nsf}
\end{figure}

\begin{figure}
\centering
\includegraphics[scale=0.33]{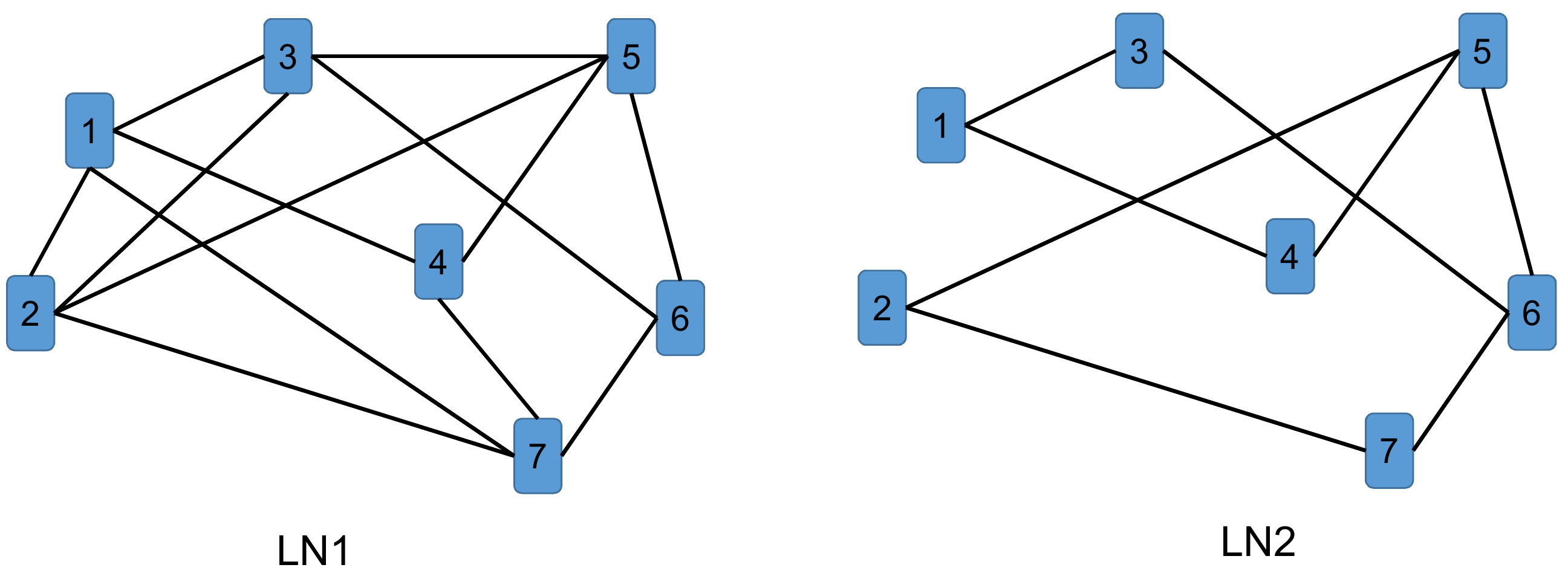}
\caption{LN1 and LN2}
\label{fig:lns}
\end{figure}

We select NSF as the physical network and create two logical networks denoted as ``LN1'' and ``LN2''. All networks are illustrated in Figs.~\ref{fig:nsf} and~\ref{fig:lns}. Two cross-layer network mappings, namely LN1-over-NSF and LN2-over-NSF, are created. We first apply the mathematical formulations in~\cite{zhou2017survivable} and verify that LN1-over-NSF is survivable and LN2-over-NSF is unsurvivable. The reason for choosing these cases is to show that our approach not only can be applied to solve both cases but also can produce results with maximal survivable probability when a survivable routing does not exist.

Physical links are assigned failure probabilities based on (1) 150 uniform failure probabilities in the range of 15\% to 0\% (decreased by 0.5\% in each step); and (2) 30 randomly generated failure probabilities with mean in the range of 15\% to 0\% and variance 2\%, where only failure probabilities above 0\% are selected. For the random failure probability, we report the average survivable probabilities of \emph{MaxPrctTree} and \emph{BasePrctTreeSet} of 5 random generated failure probability cases.

The formulations are implemented in IBM ILOG CPLEX 12.7.1 and executed on a Dell Server with 8x AMD Opteron 6366 HE Processors and 256 GB memory. Each formulation is limited to run for up to 450 seconds, and is terminated if its running time exceeds the constraint. In general, the simulations on all tested cases are completed within 1 minute.

\subsection{Simulation Results}\label{subsec:cResults}
\begin{figure}[!t]
\begin{subfigure}[b]{0.25\textwidth}
    \includegraphics[scale=0.26]{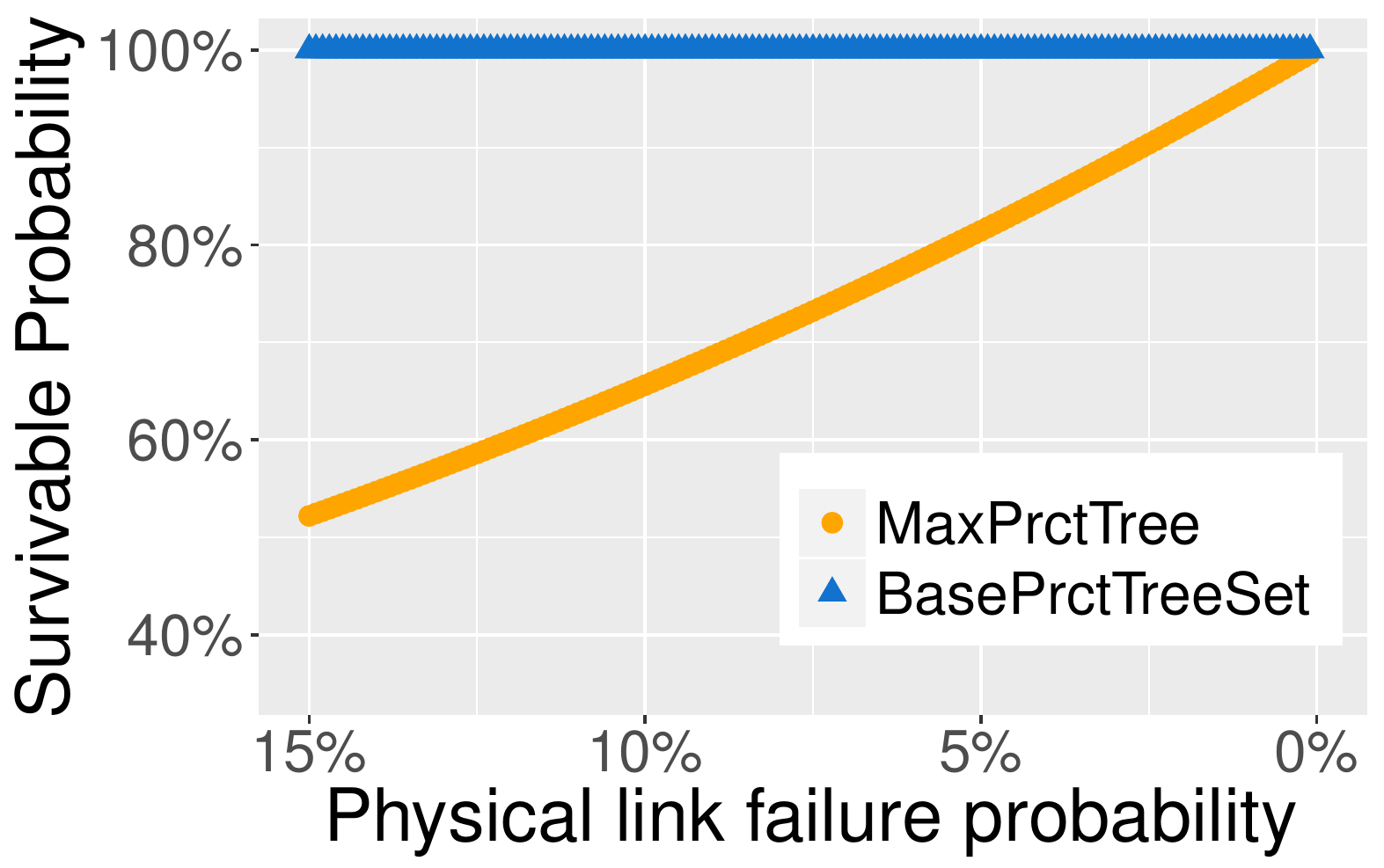}
    \caption{LN1-over-NSF}
    \label{subfig:suvUniNSF}
\end{subfigure}%
\begin{subfigure}[b]{0.25\textwidth}
    \includegraphics[scale=0.26]{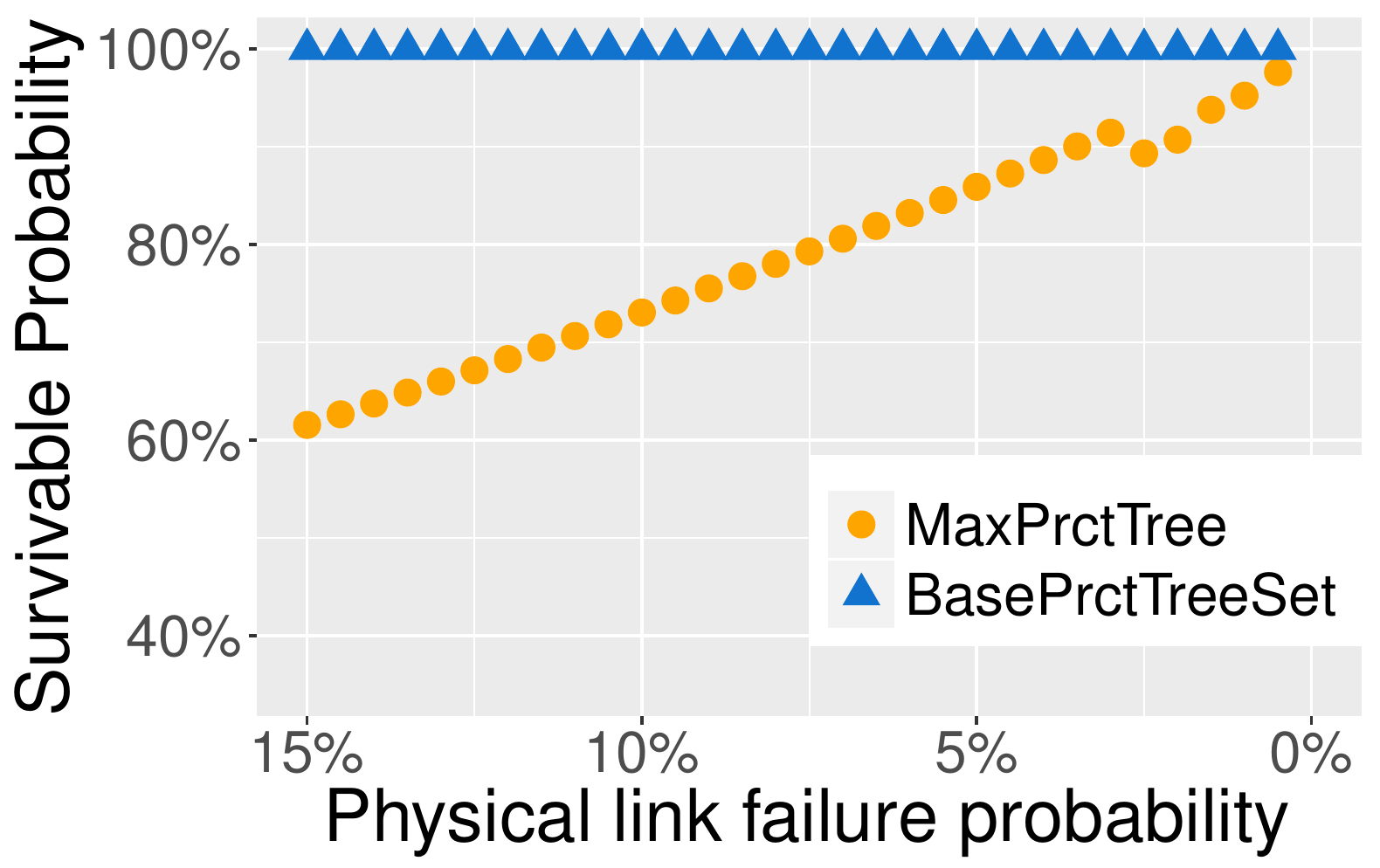}
    \caption{LN2-over-NSF}
    \label{subfig:nonsuvUniNSF}
\end{subfigure}
\caption{Survivable probability for survivable cross-layer networks}
\label{fig:survNSF}
\end{figure}

The simulations include two parts: (1) survivable cross-layer networks with LN1-over-NSF; and (2) unsurvivable cross-layer network with LN2-over-NSF with uniform and random generated physical link failure probability.
For the survivable cases, the base protecting spanning tree set is capable of providing 100\% survivability. The simulation results illustrated in Fig.~\ref{fig:survNSF} show that when the given cross-layer network is survivable, the proposed base protecting spanning tree set approach can also produce 100\% survivable design with uniform and random physical link failure probabilities.

\begin{figure}[!t]
\begin{subfigure}[b]{0.25\textwidth}
    \includegraphics[scale=0.26]{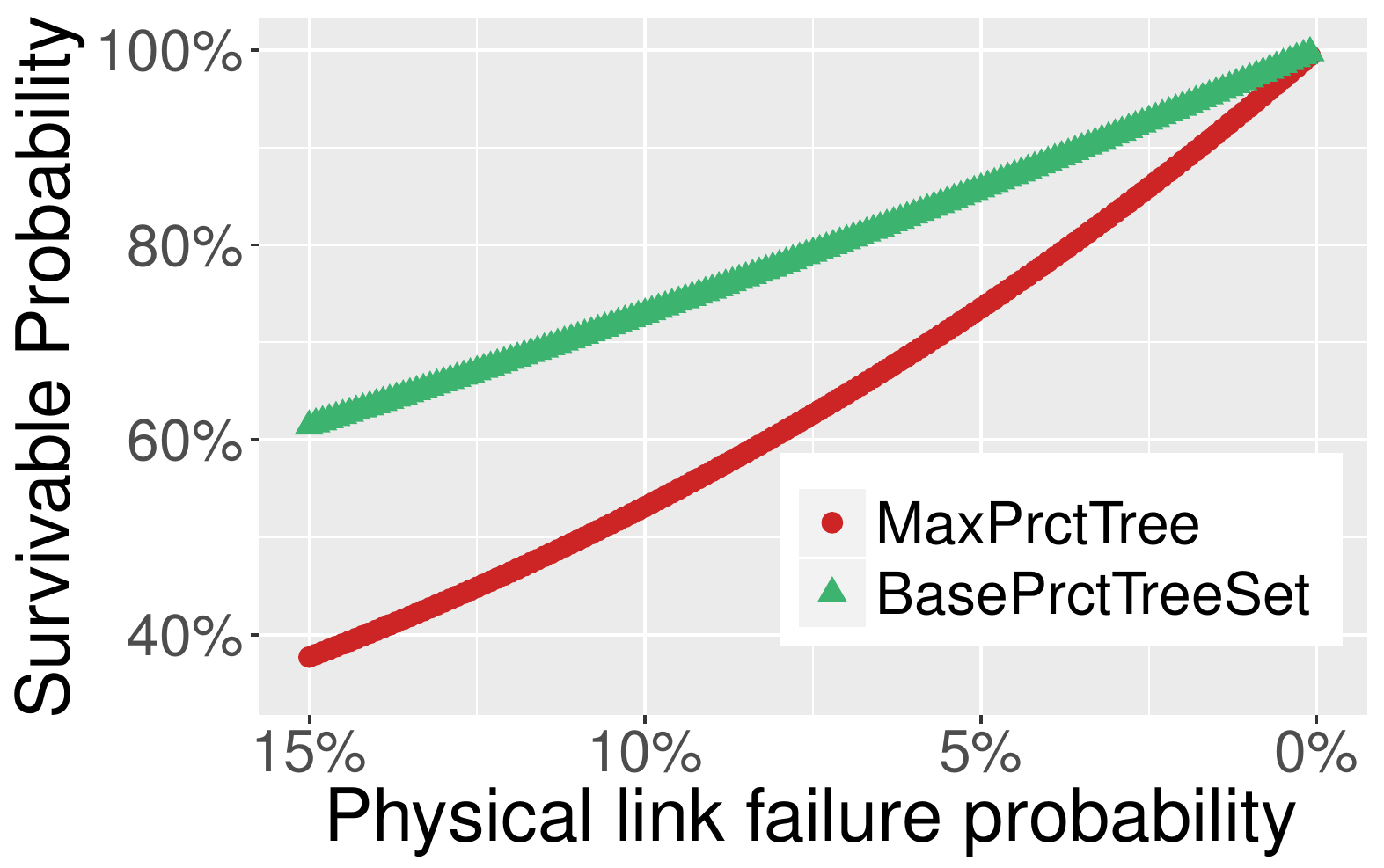}
    \caption{LN1-over-NSF}
    \label{subfig:suvUniNSF}
\end{subfigure}%
\begin{subfigure}[b]{0.25\textwidth}
    \includegraphics[scale=0.26]{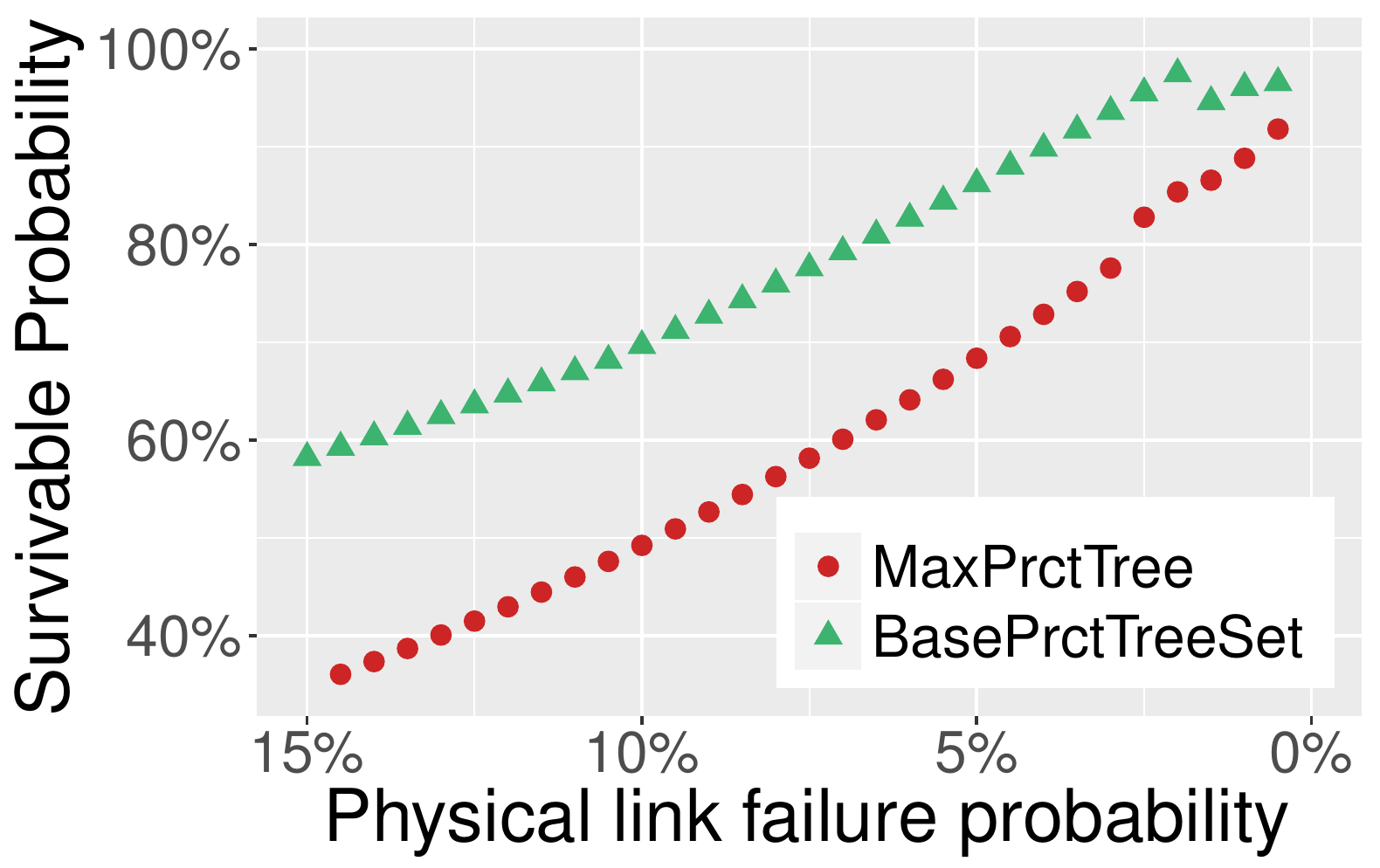}
    \caption{LN2-over-NSF}
    \label{subfig:nonsuvUniNSF}
\end{subfigure}
\caption{Survivable probability for non-survivable cross-layer networks}
\label{fig:nonSuvNSF}
\end{figure}

\begin{figure}[!t]
\begin{subfigure}[b]{0.25\textwidth}
    \includegraphics[scale=0.26]{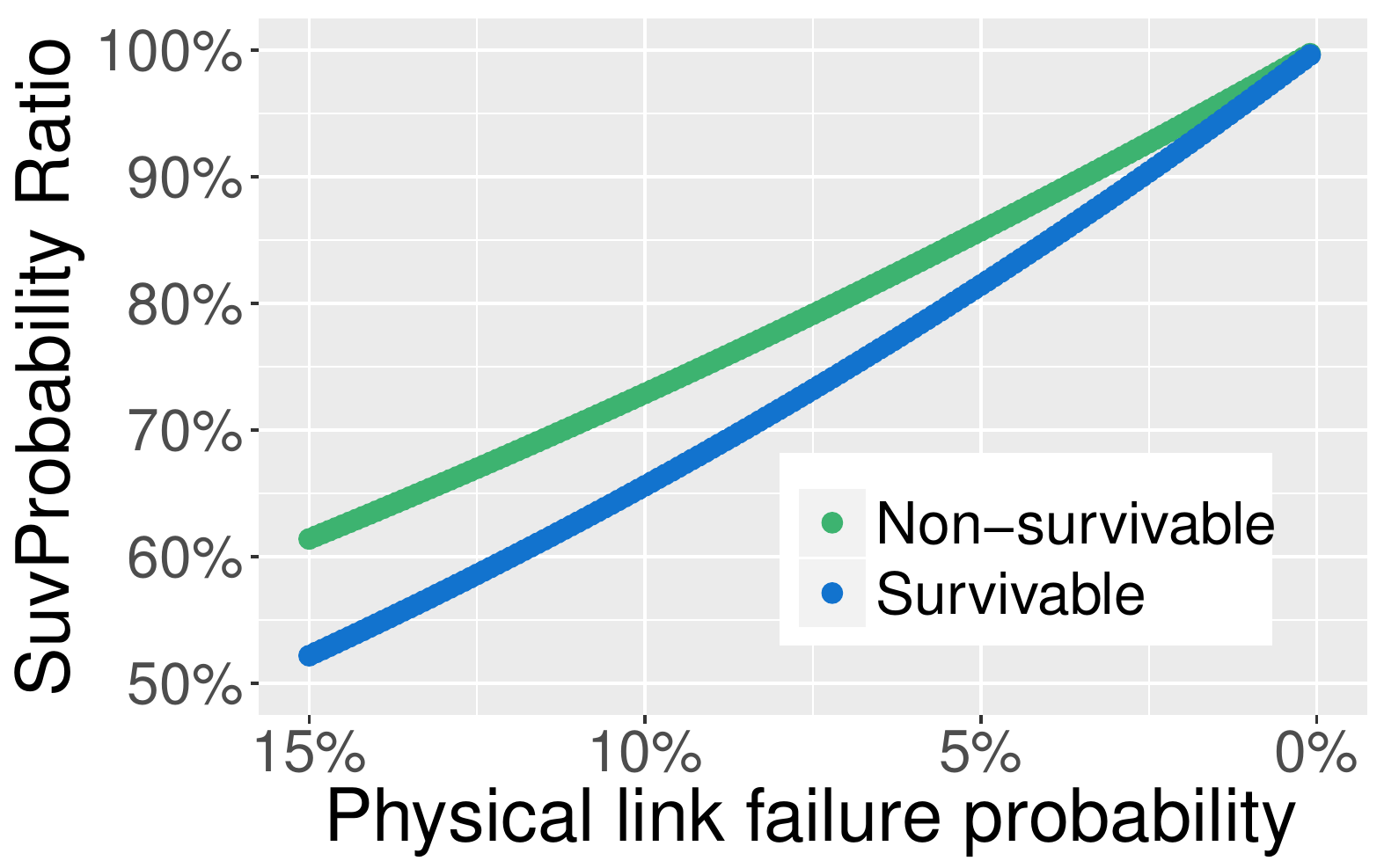}
    \caption{LN1-over-NSF}
    \label{subfig:suvUniNSF}
\end{subfigure}%
\begin{subfigure}[b]{0.25\textwidth}
    \includegraphics[scale=0.26]{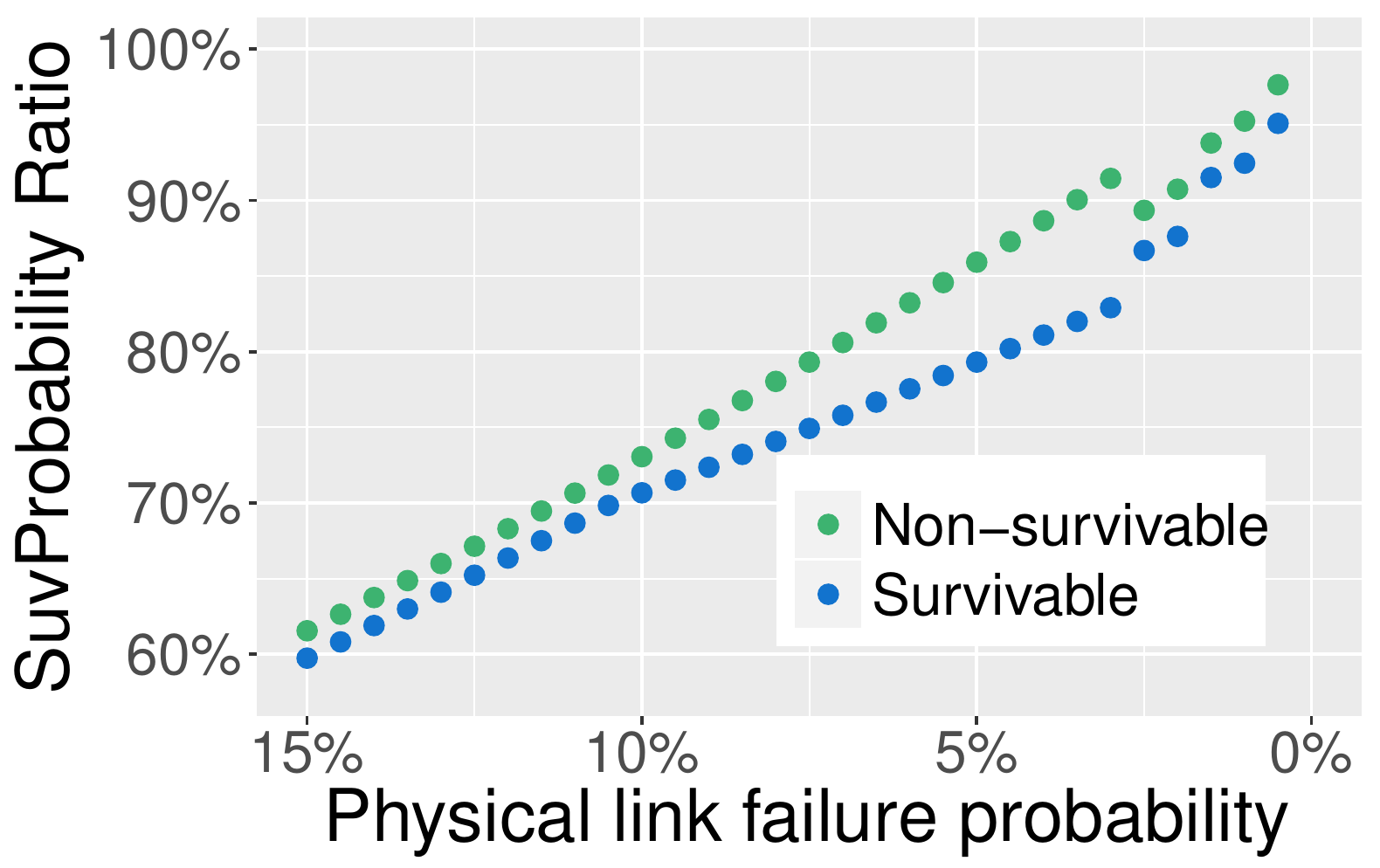}
    \caption{LN2-over-NSF}
    \label{subfig:nonsuvUniNSF}
\end{subfigure}
\caption{The ratio of survivable probability between \emph{MaxPrctTree} and \emph{BasePrctTreeSet}}
\label{fig:ratio}
\end{figure}

The survivable probability of non-survivable cross-layer network with uniform and random physical link failure probabilities are shown in Fig.~\ref{fig:nonSuvNSF}. We observe that the maximal protecting spanning tree and cross-layer survivable probability monotonically increase when the failure probability decreases.
Another interesting observation is the relation of the survivable probability between \emph{MaxPrctTree} and \emph{BasePrctTreeSet}, the maximal protecting spanning tree with a single logical protecting spanning tree providing the lower bound estimation for the survivable probability. Figure~\ref{fig:ratio} illustrates the ratio between the survivable probability of \emph{MaxPrctTree} and \emph{BasePrctTreeSet}. The results for both survivable and non-survivable networks with uniform and random physical link failure probabilities demonstrate that the lower the physical link failure probability, the better the lower bound estimation of survivable probability the maximal protecting spanning tree can provide. With up to 15\% of the average physical link failure probability, the lower bound estimation of tested cases is higher than $\frac{1}{2}$ of survivable probability of cross-layer network in all testing cases.

\section{Conclusion}\label{sec:conclusion}
In this paper, we introduced a new evaluation metric, the survivable probability, and explored exact solution approaches in the form of mathematical programming formulations. We evaluated the probability of the logical network to remain connected against physical link failure(s) with unified or random failure probabilities. We discussed the relationship of survivable probability in cross-layer networks and protecting spanning tree set, which led to the base protecting spanning tree set approach. We proved the existence of base protecting spanning tree set in a given cross-layer network and its necessary and sufficient conditions. Our simulation results show the effectiveness of proposed solution approaches.

\bibliographystyle{ieeetran}
\bibliography{IEEEabrv,pt}

\begin{thebibliography}{10}
\providecommand{\url}[1]{#1}
\csname url@samestyle\endcsname
\providecommand{\newblock}{\relax}
\providecommand{\bibinfo}[2]{#2}
\providecommand{\BIBentrySTDinterwordspacing}{\spaceskip=0pt\relax}
\providecommand{\BIBentryALTinterwordstretchfactor}{4}
\providecommand{\BIBentryALTinterwordspacing}{\spaceskip=\fontdimen2\font plus
\BIBentryALTinterwordstretchfactor\fontdimen3\font minus
  \fontdimen4\font\relax}
\providecommand{\BIBforeignlanguage}[2]{{%
\expandafter\ifx\csname l@#1\endcsname\relax
\typeout{** WARNING: IEEEtran.bst: No hyphenation pattern has been}%
\typeout{** loaded for the language `#1'. Using the pattern for}%
\typeout{** the default language instead.}%
\else
\language=\csname l@#1\endcsname
\fi
#2}}
\providecommand{\BIBdecl}{\relax}
\BIBdecl

\bibitem{develder2012optical}
C.~Develder, M.~D. Leenheer, B.~Dhoedt, M.~Pickavet, D.~Colle, F.~D. Turck, and
  P.~Demeester, ``Optical networks for grid and cloud computing applications,''
  \emph{Proc. {IEEE}}, vol. 100, no.~5, pp. 1149--1167, May 2012.

\bibitem{chowdhury2012vineyard}
\BIBentryALTinterwordspacing
M.~Chowdhury, M.~R. Rahman, and R.~Boutaba, ``{ViNEYard}: {V}irtual network
  embedding algorithms with coordinated node and link mapping,''
  \emph{{IEEE/ACM} Trans. Netw.}, vol.~20, no.~1, pp. 206--219, Feb. 2012.
  [Online]. Available: \url{http://dx.doi.org/10.1109/TNET.2011.2159308}
\BIBentrySTDinterwordspacing

\bibitem{sezer2013we}
S.~Sezer, S.~Scott-Hayward, P.~K. Chouhan, B.~Fraser, D.~Lake, J.~Finnegan,
  N.~Viljoen, M.~Miller, and N.~Rao, ``Are we ready for {SDN}? implementation
  challenges for software-defined networks,'' \emph{{IEEE} Commun. Mag.},
  vol.~51, no.~7, pp. 36--43, 2013.

\bibitem{ETSI2015NFVinfrastructure}
\BIBentryALTinterwordspacing
ETSI, ``Network functions virtualisation {(NFV)}; ecosystem; report on {SDN}
  usage in {NFV} architectural framework,'' ETSI, Tech. Rep., 2015. [Online].
  Available:
  \url{http://standards.globalspec.com/std/9979634/etsi-gs-nfv-eve-005}
\BIBentrySTDinterwordspacing

\bibitem{kang2014software}
J.-M. Kang, T.~Lin, H.~Bannazadeh, and A.~Leon-Garcia, ``Software-defined
  infrastructure and the {SAVI} testbed,'' in \emph{International Conference on
  Testbeds and Research Infrastructures}.\hskip 1em plus 0.5em minus
  0.4em\relax Springer, 2014, pp. 3--13.

\bibitem{kang2013savi}
J.-M. Kang, H.~Bannazadeh, and A.~Leon-Garcia, ``{SAVI} testbed: Control and
  management of converged virtual {ICT} resources,'' in \emph{IFIP/IEEE
  International Symposium on Integrated Network Management (IM)}.\hskip 1em
  plus 0.5em minus 0.4em\relax IEEE, 2013, pp. 664--667.

\bibitem{kreutz2015software}
D.~Kreutz, F.~M. Ramos, P.~E. Verissimo, C.~E. Rothenberg, S.~Azodolmolky, and
  S.~Uhlig, ``Software-defined networking: A comprehensive survey,''
  \emph{Proc. {IEEE}}, vol. 103, no.~1, pp. 14--76, 2015.

\bibitem{yang2015software}
M.~Yang, Y.~Li, D.~Jin, L.~Zeng, X.~Wu, and A.~V. Vasilakos, ``Software-defined
  and virtualized future mobile and wireless networks: {A} survey,''
  \emph{Mobile Networks and Applications}, vol.~20, no.~1, pp. 4--18, 2015.

\bibitem{cholda2005network}
P.~Cholda, ``Network recovery, protection and restoration of optical,
  {SONET-SDH}, {IP}, and {MPLS} [book review],'' \emph{{IEEE} Commun. Mag.},
  vol.~43, no.~7, pp. 12--12, 2005.

\bibitem{heegaard2009network}
P.~E. Heegaard and K.~S. Trivedi, ``Network survivability modeling,''
  \emph{Comput. Netw.}, vol.~53, no.~8, pp. 1215--1234, 2009.

\bibitem{ramamurthy1999survivable}
S.~Ramamurthy and B.~Mukherjee, ``Survivable {WDM} mesh networks. part
  {I}-protection,'' in \emph{Proc. IEEE INFOCOM}, vol.~2.\hskip 1em plus 0.5em
  minus 0.4em\relax IEEE, 1999, pp. 744--751.

\bibitem{KurThi05}
M.~Kurant and P.~Thiran, ``On survivable routing of mesh topologies in
  {IP}-over-{WDM} network,'' in \emph{Proc. IEEE INFOCOM}, vol.~2, 2005, pp.
  1106--1116.

\bibitem{TodRam07}
A.~Todimala and B.~Ramamurthy, ``A scalable approach for survivable virtual
  topology routing in optical {WDM} networks,'' \emph{{IEEE} J. Sel. Areas
  Commun.}, vol.~25, pp. 63--69, 2007.

\bibitem{parandehgheibi2014survivable}
M.~Parandehgheibi, H.-W. Lee, and E.~Modiano, ``Survivable path sets: {A} new
  approach to survivability in multilayer networks,'' \emph{J. Lightw.
  Technol.}, vol.~32, no.~24, pp. 4139--4150, 2014.

\bibitem{ModNar01}
E.~Modiano and A.~Narula-Tam, ``Survivable routing of logical topologies in
  {WDM} networks,'' in \emph{Proc. IEEE INFOCOM}, vol.~1, 2001, pp. 348--357.

\bibitem{lee2011cross}
K.~Lee, E.~Modiano, and H.-W. Lee, ``Cross-layer survivability in {WDM}-based
  networks,'' \emph{{IEEE/ACM} Trans. Netw.}, vol.~19, no.~4, pp. 1000--1013,
  2011.

\bibitem{rahman2013svne}
M.~R. Rahman and R.~Boutaba, ``{SVNE}: {S}urvivable virtual network embedding
  algorithms for network virtualization,'' \emph{{IEEE} Trans. Netw. Service
  Manag.}, vol.~10, no.~2, pp. 105--118, 2013.

\bibitem{ye2014survivable}
Z.~Ye, A.~N. Patel, P.~N. Ji, and C.~Qiao, ``Survivable virtual infrastructure
  mapping over transport software-defined networks {(T-SDN)},'' in
  \emph{Optical Fiber Communications Conference and Exhibition (OFC)}.\hskip
  1em plus 0.5em minus 0.4em\relax IEEE, 2014, pp. 1--3.

\bibitem{ye2015survivable}
------, ``Survivable virtual infrastructure mapping with dedicated protection
  in transport software-defined networks,'' \emph{{IEEE} J. Opt. Commun.
  Netw.}, vol.~7, no.~2, pp. A183--A189, 2015.

\bibitem{zhou2017survivable}
Z.~Zhou, T.~Lin, and K.~Thulasiraman, ``Survivable cloud network design against
  multiple failures through protecting spanning trees,'' \emph{J. Lightw.
  Technol.}, vol.~35, no.~2, pp. 288--298, 2017.

\bibitem{zhou2017novel}
Z.~Zhou, T.~Lin, K.~Thulasiraman, and G.~Xue, ``Novel survivable logical
  topology routing by logical protecting spanning trees in {IP-over-WDM}
  networks,'' \emph{{IEEE/ACM} Trans. Netw.}, no.~3, pp. 1673--1685, 2017.

\bibitem{yallouz2014tunable}
J.~Yallouz, O.~Rottenstreich, and A.~Orda, ``Tunable survivable spanning
  trees,'' \emph{ACM SIGMETRICS Performance Evaluation Review}, vol.~42, no.~1,
  pp. 315--327, 2014.

\bibitem{yallouz2017tunable}
J.~Yallouz and A.~Orda, ``Tunable {QoS}-aware network survivability,''
  \emph{{IEEE/ACM} Trans. Netw.}, vol.~25, no.~1, pp. 139--149, 2017.

\bibitem{zhou2012}
Z.~Zhou, T.~Lin, K.~Thulasiraman, G.~Xue, and S.~Sahni, ``Novel survivable
  logical topology routing in {IP}-over-{WDM} networks by logical protecting
  spanning tree set,'' in \emph{Proc. Intl. Congress on Ultra Modern Telecomm.
  and Control Sys.}, Oct 2012, pp. 650--656.

\bibitem{lee2014maximizing}
H.-W. Lee, K.~Lee, and E.~Modiano, ``Maximizing reliability in {WDM} networks
  through lightpath routing,'' \emph{{IEEE/ACM} Trans. Netw.}, vol.~22, no.~4,
  pp. 1052--1066, 2014.

\end{thebibliography}

\end{document}